\documentclass[12pt]{article}
\usepackage{amsmath}
\usepackage{graphicx}
\begin{document}
\title{Putting Together A Cyclical Baryonic Universe}
\author{David E. Rosenberg,  \\LSU Relativity Group  Loni Hyrel 17, Baton Rouge LA 70808 USA \\copyright\,{2018}\\ \small{email: drosenberg@hrchc.org, tel. 757-393-6363, fax 757-397-0047}} 
\maketitle
\begin{abstract}
There are multiple examples of unexplained gravitational losses in neutron stars and black holes.
Protons and neutrons have been found to have enormous repulsive pressures that neutron stars and even collapsing black holes (highly squeezed matter) can not overcome. The case against singularities follows. Galactic black hole gravitational loss can supply the missing dark energy. With highly squeezed nucleons, the Big Bang could begin with a hot core and cold dark matter shell. The 21 cm. radiation data has identified baryon sized paticles as cold dark matter. Highly squeezed nucleons do not decompose to produce anti-matter. The CBR Planck Spectrum was generated by core radiation and highly conducting walls. The flatness of the Universe is due to a bounce. The highly correlated galaxies originated from primordial black holes  capturing and sequestering hot core gases in low orbits. There is evidence that galaxies have not grown nor merged significantly since formation. 
\end{abstract}
\newpage
\small
   This paper tying the universe together is arranged as follows:

1. Unexplained gravitation loss.................................................... ..................2

2. The case against singularities......................................................................5

3. Explaining gravitation losses.......................................................................5

4. A Cyclical Universe.....................................................................................12

5. Current problems in galaxy formation.........................................................16

6. Baryonic galaxy formation..........................................................................18

7. Discussion...................................................................................................22

\section{Unexplained Gravitation Loses}

General relativity has been most successful in explaining the universe except in extreme density conditions: the big bang and black holes. Galaxies are constructed similarly despite their origins being physically too fair apart to be in casual contact (the horizon problem). Initial spacetime was Minkowskian-the expansion energy exactly matched the gravitational energy (the flatness problem). The hot synthesis of light elements occurred in only $4-5\%$ of the matter present. Extrapolation of general relativity in the big bang and black holes to or near singularities has not been successful in solving the universe problems. There is no evidence that the universe ever reached Planck energies. In the 'Road to Relativity' Einstein felt the gravitational tensors were 'rock solid' but stress-energy tensors were made of 'wood'.      

Neutron stars have crusts, a surface region with densities  $< 10^{14} g/cm^{3}$. Here in beta equilibrium, there are neutrons, nuclei and electrons. Relativistic degenerate electrons comprise most of the pressure. The baryon density is near nuclear saturation density $n_{0}\approx 0.16\,fm^{3}$. The general presumption has been that the core should be contain even more relativistic gravitational matter.  
Applying general relativity to a static spherical symmetric metric gives
\begin{equation}
ds^{2}=-e^{2\Phi(r)}c^{2}dt^{2}+e^{2\Lambda(r)}dr^{2}+r^{2}d\theta^{2}+r^{2}sin^(2)\theta\, d\phi^{2}
\end{equation}
where $\phi$ is the azimuthal angle, $\theta$ is the polar angle and the radial coordinate $r$ is defined such that  
at the origin the circumference of a circle is $2\pi r$. Using a Schwarzschild at the star surface, the boundary condition mandates 
\begin{equation}
\Phi(r=R)=\frac{1}{2}ln \Bigl(1-\frac{2GM}{Rc^{2}}\Bigr)
\end{equation}
\begin{equation}
\Lambda(r=R)=-\frac{1}{2}ln\Bigl(1-\frac{2GM}{Rc^{2}}\Bigr)
\end{equation}
If one assumes a perfect fluid to oversimplify calculations, this leads to the Tolman-Oppenheimer-Volkov (TOV) equations:
\begin{equation}
\frac{dP}{dr}=-\frac{Gm\rho}{r^{2}}\Bigl(1+\frac{P}{\rho c^{2}}\Bigr)\Bigl(1+\frac{4\pi r^{3}P}{mc^{2}}\Bigr) \Bigl(
1-\frac{2Gm}{rc^{2}}\Bigr)^{-1}
\end{equation}
\begin{equation}
\frac{dm}{dr}=4\pi r^{2} \rho
\end{equation}
Here $P=P(r)$, $\rho=\rho(r)$ and the mass wihin the radius $r$ is $m(r)$. In that the TOV equations are relativistic, the pressure $P$ should also add to gravitation especially at the core. At this boundary condition,
$m(0)=0$, $\rho(0)=\rho_{c}$. With general relativity the pressure terms in the TOV equations will add much gravitation. It has been found that squeezing protons to 0.3 fm (femtometer) yields an enormous resistive pressure of $10^{35}$ pascals\cite{Burkert et al. (2018)}. As they note, this is more pressure than in a neutron star core. The following examples are going to show that gravitation is reduced both in neutron star cores and black holes. We will then explain how highly squeezed nucleons will reduce or eliminate gravitation.   

In neutron stars: There is the unreasonable effectiveness of the post-Newtonian approximation in strong gravitational fields of neutron stars\cite{Will (2011)}. This approximation assumes that gravitational fields in and around bodies are weak and the motions of matter are slow compared to the speed of light. Thus
\begin{equation}
(v/c)^{2}\sim GM/\tau c^{2} \sim p/\rho c^{2} << 0.1
\end{equation}
where $v, M \,and\, \tau$ are velocity, mass and separation of the system. Within masses $p$ and $\rho$ are the pressure and density and $G \, and \, c$ are Newton's gravitational constant and the speed of light. What is used in post-Newtonian calculation is the 'reduced' Einstein equation
\begin{equation}
(-\partial^{2}/\partial(ct)^{2}+\nabla^{2})h^{\alpha\beta}=-16\pi(G/c^{4})
\tau^{\alpha\beta}
\end{equation}
where $h^{\alpha\beta}$ is the deviation of the space-time metric $g_{\alpha\beta}$ from the flat Minkowski space-time metric $\eta_{\alpha\beta}$. They are related by the equation $h^{\alpha\beta}\equiv\eta^{\alpha\beta}-(-g)^{1/2}g^{\alpha\beta}$. If gravity is weak everywhere, then the gravitational tensor potential $h^{\alpha\beta}$ must be small and can approximate the  highly nonlinear equations of general relativity.

In 1974 the binary pulsar PSR \,$1913+16$ was discovered. Each neutron star had a mass $\approx 1.4M_\odot$ in a quite relativistic orbital system with a mean speed $\approx 200 \, km/sec$. Strangely, it was found that the rate of decay of the orbit was in agreement with the post-Newtonian quadripole formula. More recently, the relativistic double pulsar $J \, 0737-3039$ performed according to the post-Newtonian calculations despite being in very strong gravitational fields. An effacement process had been postulated that the gravitational binding energy reduces the gravitational mass of each pulsar by $10-20\%$ compared to its rest mass.  

A low mass Xray binary AXJ1745.62901 has steadily been losing extra orbital momentum over a 20 year period\cite{Ponti et al. (2015)}. The high rate of orbital period decrease is $\dot{P}_{orb}= -4.03\pm 0.32 \times 10^{-11} s/s$. This is over an order of magnitude greater than expected loses due to gravitational waves, magnetic breaking, or conservative mass transfer. The data included unexplained 'jitter' of $10-20$ seconds advancing or retarding the orbital period. Orbital loss explanations from accretion, a third body mass or an unrealistic $.001M_{\odot}$ in the outer disc were rejected.

Looking at neutron stars and black hole formation, there is an  gap between the largest neutron star of $1.97M_\odot$ and the smallest black holes $>4M_\odot$\cite{Kreidberg et al. (2012)}.
Black holes begin to form at densities
\begin{equation}
\rho = (c^6)/(G^3M^2)
\end{equation}
Since the constants $c^{6}/G^{3} =6.272 \times 10^{84} grams^{3}/cm^{3}$, $\sim 10^{9} M_{\odot}$ of water will collapse to a black hole. The smallest black hole of $4M_{\odot}$ will form at highest densities.

There is the variation of Newton's Gravitational constant G with the length of day 
\cite{Anderson et al. (2015)}. 
G is integral part of Newton's gravitational law between two massive bodies: $Force=Gm_{1}m_{2}/r^{2}$ as well as the General Relativity of Einstein $R_{ab}-(1/2)\, g_{ab}R=8\pi\, G\,T_{ab} $
where $m_{1}$ and $m_{2}$ are massive bodies and $r$ is the separation distance between. $R_{ab}$ and $R$ are the Ricci tensor and scalar, respectively and $T_{ab}$ is the stress-energy tensor. They analyzed $13$ measurements of $G$ and found that it was greater than 
$6.674 \times 10^{-11} m^{3}kg^{-1}s^{-2}$ when the days were longer (more solar radiation). It was less than when the days were shorter. This change in gravitation seems related to the increases or decreases in kinetic energy of particles.   

Our galaxy shows evidence of energy and gravitational loss from the super-massive black hole after billions of years\cite{Blitz (2011)}. There is a lopsided thickness in the outer disk of the Milky Way. This is about twice as large in one side as the other, despite the spherically symmetric distribution of galactic dark matter. 

There is the mysterious survivor of an encounter with our Milky Way Super-massive Black Hole\cite{Bally (2015)}. As far back as 2006 a G2 'cloud' was found accelerating toward our galactic center mass  $\approx 3\times 10^{6}\,M_\odot$. The closest approach took place about February 2014. By Sept. 2014 G2 cloud was mysteriously moving away from the black hole. It even failed to trigger a flare up of accretion activity. 

There is the amazingly weak magnetic field of a $9 M_{\odot}$ black hole in the binary V404 Cygni system\cite{Dallilar et al. (2017)}. A burst of radiation was studied from a flare. It contained charged particles with electrons and protons in the black hole magnetic field. By calculating how quickly the burst dimmed, a team of astronomers found the magnetic field to be $461\pm 12$ gauss, the strength of several bar magnets and almost 3 orders of magnitude less than theory predicted.

\section{The Case Against Singularities}

Supposedly physical infinities are the singularities found in black holes and where the big bang started. 
If the universe started at or near a singularity and expanded, there should remain evidence of many of the following missing high energy phenomena. Monopoles are formed $>10^{14}$ GeV. Antimatter is formed $>38$ MeV. Domain walls would be the size of $10^{28}h^{-1}$ cm. and have a mass $\sim 4 \times 10^{65}\Lambda^{1/2}(\sigma/100 GeV)^{3}$ grams or many orders of magnitude over the present Hubble volume. Its prescence would cause a large defect in the CBR.
Assuming the universe started as a fireball, the production of geometric flatness $\Bigl(a_{,t}/a\Bigr)^{2}=-k/a^{2}+\Lambda/3+\, 8\pi\rho_{m}a_{o}^{3}/3a^{3}+\,8\pi\rho_{r}a_{o}^{4}/3a^{4}$ with $k=0$ has required an unproven inflaton scalar $\phi$, leaving most other problems unsolved. Fireballs can't make high correlations in galaxies with the velocity-brightness relation, corresponding rotations and all galactic parameters related to the central black hole size(see below).   
Considering the small size of the Higgs Boson (125 GeV), the big bang could not produce any cold dark nonbaryonic matter unless strange quarks were involved. Nucleosynthesis (the highest big bang energy confirmed) occurred after the end of quark-hadron boundary, requiring $\approx 95\%$ cold dark matter. 

This paper has shown examples of black holes and neutron stars losing gravitation. What kind of black hole singularity can lose gravitational energy? Where are all the black holes of $1,2,3 M_{\odot}$? Why can't large collapsing stars reach the densities to make them? They certainly won't have evaporated like microscopic black holes. 
How can neutron stars lose gravitational energy when their cores are under $8\times$ nuclear density? Why are neutron stars limited to $2M_{\odot}$? Why is the range $2-3M_{\odot}$ unstable? It seems the formation of $4M_{\odot}$ black holes are the highest density that can be reached. 
 Neutron stars have a mobile non-gravitational core which is heavy enough to cause a 'wobble' in the orbit.  
 
\section{Expaining Gravitation Loses}
Gravitational masses may be treated as pinpoint sources. However, it defies logic that the gravitational properties of an infinitely collapsing mass would be the same as normal matter.  
For gravitational losses, matter must be highly squeezed and not a gas of noninteracting particles (perfect fluid) nor a Quark-Gluon Plasma. Particles must be packed so tightly by $10^{16}gm/cm^{3}$ that there is no motion at all.

On constructing the stress-energy tensor\cite{Misner et al. (1973)}, the matter 4-velocity $\mathbf{u}=(dt,0,0,0)$, possibly adding only the observer 4-velocity. The 4-momentum is
\begin{equation}
\mathbf{p} = m\mathbf{u}=(m\gamma, m\Delta x \gamma, m\Delta y \gamma, m\Delta z \gamma)  
\end{equation}
where $E=m\gamma$, $\gamma =1$ nonrelativistic and $\Delta x, \Delta y, \Delta z =0$. A volume element $\Sigma$ composed of basis vectors $e_{x},e_{y}, e_{z}$ have zero magnitudes as there is no flow of 4-momentum.   
An observer in his Lorentz frame will measure mass-energy density in $gm/cm^{3} \, T_{00}=T(e_{0},e_{0})$, with the observers 4-velocity $\mathbf{u}$ replaced by $e_{0}$.  
If the box is at rest in the observers frame, all matter kinetic and quantum energy will be sufficiently damped except possibly quantum spin. Space-time interaction will cease, $T_{00}=0$.
To construct a volume in spacetime with a parallelopipid, use four different vectors for edges $\mathbf{A,B,C,D}$. The vectors in standard Lorentz frame are $\mathbf{A}=(\Delta t, 0, 0, 0)$, $\mathbf{B}=(0, \Delta x, 0, 0)$, $\mathbf{C}=(0, 0, \Delta y, 0)$ and $\mathbf{D}=(0, 0, 0, \Delta z)$. 
A 4-volume is
\begin{equation}
\Omega=\epsilon_{\alpha \beta \gamma \delta}A^{\alpha}B^{\beta}C^{\gamma}D^{\delta}=\mathbf{A} \wedge \mathbf{B} \wedge \mathbf{C} \wedge \mathbf{D}
\end{equation}
A volume integral of a tensor \textbf{S} defined over a four dimensional  region $\mathcal{V}$ of spacetime, calculated in a Lorentz frame 
\begin{equation}
M^{\alpha}_{\beta \alpha}=\int S^{\alpha}_{\beta \gamma} dt\, dx \, dy \, dz
\end{equation} 
 The energy density measured in such a volume $E=m\gamma/V=0$  as is the density of the 4-momentum $d\mathbf{p}/dV=0$ per 3 dimentional volume in an observers Lorentz frame.  In the following, $j,k=(1,2, or \,3)$ in what really is a symmetric tensor. $T^{j,0}=0$ is the momentum density, j component. $T^{0,k}=0$ is the energy flux, k component. $T^{j,k}=0$ is the j component of force from matter and fields acting around $x^{k}$.
This keeps the tensor divergence $\mathbf{\nabla} \cdot \mathbf{T}=0$ as nothing is moving.

For rotations of immobile squeezed nucleons, let S be a spacelike hypersuface with arbitrary event $\mathcal{A}$ and coordinates $x^{\alpha}(\mathcal{A})\equiv a^{\alpha}$ using globally inertial coordinates.  
Total angular momentum  on S about $\mathcal{A}$ can be defined as 
\begin{equation}
J^{\mu\nu} \equiv \int_{S} \mathcal{J}^{\mu\nu\alpha} d^{3}\sum_{\alpha}
\end{equation} and will add to total momentum only if present. 
Here 
\begin{equation}
\mathcal{J}^{ \mu\nu\alpha}\equiv(x^{\mu}-a^{\mu})T^{\nu\alpha}-(x^{\nu}-\mathcal{A}^{\nu})T^{\mu\alpha}
\end{equation}
If S is a hypersurface of constant time t then 
\begin{equation}
J^{\mu\nu}=\int \mathcal{J}^{\mu\nu 0} dx\, dy\, dz
\end{equation}
In the systems rest frame,  let $P^{0}=M$, $P^{j}=0$ and at the center of mass 
\begin{equation}
x_{cm}^{j}=\frac{1}{M}\int x^{j}T^{00}d^{3}x.
\end{equation}
For a large mass, intrinsic angular momentum  is defined angular momentum about any event $(a^{0},x_{cm}^{j})$ on the world line of the center of mass. 
Here components $S^{0j}=0$ and  
\begin{equation}
 S^{jk}=\epsilon^{jkl}S^{l}.\mbox{ and }  S\equiv \int (x-x_{cm})\times d\mathbf{p}/dV \,S^{\mu\nu} d^{3}x
\end{equation} 
The intrinsic angular momentum 4-vector $S^{\mu}$ has components in the rest frame (0,S) 
\begin{equation}
S^{\mu\nu}=U_{\alpha}S_{\beta}\epsilon^{\alpha\beta\mu\nu}
\end{equation} 
The 4-velocity center of a large highly squeezed mass $\mathbf{U}_{\beta}\equiv \mathbf{P}_{\beta}/M =0$.   
Angular momentum is composed of intrinsic and orbital parts. 
An arbitrary event $a$ whose perpendicular distance from the center of mass world line is $-Y^{\alpha}$ making $\mathbf{U}_{\beta}Y^{\beta}=0$.
The total angular momentum $J^{\mu\nu}$ about $\mathcal{A}$ is both the intrinsic part 
\begin{equation}
(S^{\mu\nu}=\mathbf{U}_{\alpha}\mathbf{S}_{\beta}\epsilon^{\alpha\beta\mu\nu})
\end{equation}
 and the orbital part 
\begin{equation}
(L^{\mu\nu}=Y^{\mu} P^{\nu}-Y^{\nu} P^{\mu})
\end{equation}
  With the angular momentum about $\mathcal{A}$ and the 4-momentum known (zero is this case) one can calculate
the vector from $\mathcal{A}$ to the center of mass world line.
\begin{equation}
Y^{\mu}=-J^{\mu\nu}P_{\nu}/M^{2}
\end{equation}
 
In a swarm of identical particles with event $\mathcal{P}$ inside the swarm, $m_{A}$ is the rest mass. $\mathbf{u}_{A}$ is the 4-velocity, and $\mathbf{p}_{A}$ is the 4-momentum. 
$N_{A}$ is the number of particles per unit volume, as measured in the particles own rest frame. 
The number flux vector 
\begin{equation}
\mathbf{S}_{A}\equiv N_{A}\mathbf{u}_{A}
\end{equation}
 The particles have ordinary velocity $v_{A}$, zero in packed supranuclear densities. 
$\mathbf{u}^{o}_{A}$  is the the Lorentz correction for volume and velocity  $1/(1-v_{A})^{1/2}$. 
The 4-momentum density  is 
\begin{equation}
\mathbf{p}_{A} S^{o}_{A}=m_{A}u^{u}N_{A}u^{o}_{A}
\end{equation} 
Consequently the 4-momentum density has components 
\begin{equation}
T^{uo}_{A}= p^{u}_{A}S^{o}_{A}=m_{A}N_{A}u^{u}_{A}u^{o}_{A}
\end{equation} 
The flux of the $\mu$ component of momentum with perpendicular projection $e_{j}$ is 
\begin{equation}
 T^{\mu j}_{A}= p^{\mu}_{A}S^{j}_{A}=   m_{A}N_{A}u^{\mu}_{A}u^{j}_{A}
\end{equation} 
Here suprascripts $(\mu,o)$ and $(\mu,j)$ of the frame independent equation
\begin{equation}
\mathbf{T}_{A}= m_{A}N_{A}\mathbf{u}_{A} \otimes \mathbf{u}_{A}= \mathbf{p}_{A} \otimes \mathbf{S}_{A}
\end{equation}
By summing over all categories, the total number flux vector and stress energy tensor are obtained for all particles in the swarm. If $\mathbf{u}_{A} =0$, these will be zero. 
\begin{equation}
\mathbf{S}=\sum_{A}N_{A}\mathbf{u}_{A} \mbox{ and }\mathbf{T}=\sum_{A} m_{A}N_{A}\mathbf{u}_{A} \otimes\mathbf{u}_{A}=\sum_{A}\mathbf{p}_{A}\otimes\mathbf{S}_{A}
\end{equation} 
The total momentum flux accross a closed 3-dimensional surface must vanish $\oint T^{\mu 0} d^{3}\sigma_{a} =0$. 
There is no flux and no sinks and there is no momentum at these supranuclear densities. 

Thorne's hoop conjecture states a black hole forms only when a given amount of mass-energy collapses through its own Schwarzschild radius $R_{S}=2M$, thus achieving compactness $M/R>0.5$. A sufficient large collapsing shell is necessary. If one starts with collapsing black hole nucleons maximally squeezed to $0.3$ fm radius (each with internal repulsive pressures $\sim 10^{35}$ pascals and probably much more together), they can not be packed to more than $\approx 10^{16}gm/cm^{3}$ density. As noted above, due to complete core gravitational loss, collapsing matter can not overcome these internal pressures. This is the core of black holes. Not near Planck energies. Like neutron star cores, black holes will begin to lose gravitation by $\approx 5 \times 10^{14} gm/cm^{3}$. Once the core begins to form, its little to no gravitational energy and virtual incompressiblity blocks further gravitational collapse. It takes a total of $4M_{\odot}$, including the non-gravitational core to produce a black hole. The presence of a nongravitational core will reduce $M_{gravitation}$ but otherwise not affect gravitational waves in black hole-black hole coalescence. The numerical Einstein equations have been found to match the data in black hole-black hole coalescence GW 150914 \cite{Abbott et al. (2016)}. Black hole gravitational loss would not be evident unless there was a binary black hole system. This constrains semianalytical models, two independent precessing spins and spin weighted quadripole modes. Component spins in magnitude and direction were not constrained. Using 
the conformally related metric $\bar{\gamma}_{ij}$ and the conformal factor $e^{\phi}$, the Hamiltonian constraint is
\begin{equation}
0=\mathcal{H}=\bar{\gamma}^{ij} \bar{D}_{i}\bar{D}_{j}e^{\phi}-e^{\phi}/8\bar{R}+e^{5\phi}/8\tilde{A}_{ij}\tilde{A}^{ij} e^{5\phi}/12 K^{2} 2\pi e^{5\phi}\rho
\end{equation}
The only other source equation is a momentum constraint with precessing spin $S^{i}=\rho h Wu_{i}$ is the momentum density.
\begin{equation}
0=\mathcal{M}=\bar{D}_{j}(e^{6\phi}\tilde{A}^{ji})-2/3e^{6\phi}\bar{D}^{j}K-8\pi e^{6\phi}S^{i}
\end{equation}
Therefore these equations will start to fail at neutron star core densities. All singularities are mathematical but not physical. Black holes  will collapse until $G_{\mu\nu}\rightarrow \mathbf{0} $. Likewise, the prior universe matter (pre-big bang) collapsed till $G_{\mu\nu}\rightarrow \mathbf{0} $. For a black hole in Schwarzschild geometry,
\begin{equation}
ds^{2}= \Bigl(1-M/r\Bigr)dt^{2}+\Bigl(1-2M/r\Bigr)^{-1}dr^{2}+r^{2}d\theta^{2}+r^{2}\, sin^{2}\theta\, d\phi^{2}
\end{equation}
where $M_{gravitational}\rightarrow 0$ very slowly, more so for the largest masses. 

In neutron stars, gravitational losses must begin in their cores with density ranges $10^{14-15} gm/cm^{3}$. With a normal stressed medium having velocities $|\mathbf{v}|<< 1$ with respect to a specific Lorentz frame, the spacial components of the momentum are $T^{0j}=\sum m^{jk} v^{k} $. Here 
\begin{equation}
m^{jk}= T^{\bar{0}\bar{0}}\delta^{jk}+ T^{\bar{j}\bar{k}}
\end{equation} 
Here $T^{\bar{\mu}\bar{\nu}}$ are components of the stress energy tensor in the rest frame of the medium. 
Inside a neutron star core, where velocities are known low, and probably zero $T^{\bar{0}\bar{0}} =0 \sim T^{\bar{j}\bar{k}}$. 

After a collapse is halted, the most highly squeezed particles in the core will act as an energy sink. 
Geometrized units will be employed such that $c=1=G$. 
From the second law of thermodynamics
\begin{equation}
T\, ds=d(\rho V) + pdV=d[V(\rho + p)]-Vdp
\end{equation}
here $s$ is the entropy of the matter and $V \propto a^{3}$ is the co-moving volume. 
Integrating gives $dp=(\rho + p)/T \, dT$. Substituting this into the above gives two equations,
\begin{equation}
d/dt\Bigl[(V(\rho + p))/(T)\Bigr]=(VE_{s})/(T)
\end{equation} 
and
\begin{equation}
\int ds= \int d\Bigl[ (V(\rho + p))/(T) \Bigr]
\end{equation}
The shell entropy change is $\dot{s}_{2}=E_{s}V/T_{2}$ and the core is $\dot{s_{1}}= - E_{s}V/T_{1}$.
Normally the energy and entropy increment would follow the temperature differential to a lower temperature as
with radiation and dust or with a scalar field and radiation.
\begin{equation}
\dot{s}_{total}=\dot{s_{2}} + \dot{s_{1}}=E_{s}V\Bigl(1/T_{2} -1/T_{1} \Bigr)
\end{equation}
Quantum effects cause some strange properties. 
Temperature is a function 
of the velocity or kinetic energies of the particles, $T^{o}K=m\bar{v}^{2}/3$. 
\begin {equation}
\int d\rho=\int ((\rho+p)/(n)\,dn-nTds)
\end{equation}
Thus the core temperature and entropy will be reduced in highly squeezed particles by gravitational caused confinement. 
Due to quantum gravity effects, core matter, with the most highly squeezed particles,
will act like an continuous energy sink and drain surrounding shell matter of most energies. Thus cold dark matter originated in shell baryons. The big bang hot core was about $4-5\%$ of the total mass.
There is still baryon conservation $dn/d\tau =-n\nabla \cdot u$. Energy conservation will still occur.
Only $d(nsV)/d\tau \leq 0$. Quantum gravity effects will put matter in better order, that is with less entropy and kinetic energy. These effects will also solve the following:

Between $>2M_\odot$ and $<4M_{\odot}$ neutron stars must be unstable due to noncollapsing cores causing collapsing matter to bounce (explode). It has been postulated that the bouncing will occur when  inner core mass 
\begin{equation}
M_{IC}\propto Y_{e}^{2}(1+\eta S^{2}) 
\end{equation}
 has the specific entropy $S \approx 3 $. Here $\eta =0.1$ and $Y_{e}$ the electron/baryon ratio. The quantities are related through electron capture, neutrino transport and equalization of trapped neutrinos. Core collapse separates a star into subsonic  collapsing inner core and supersonic outer core regions. The prompt shock fails to blow up the star and stalls due to neutrino energy losses and incoming collapsing matter. Despite neutrino heating of collapsing layers, violent sloshing motions and rotations-known as standing accretion shock instability, explosions are uncertain even on the biggest computers. A virtually incompressible inner core of highly squeezed nucleons is able to solve the long standing problem of shock revival with a bounce causing a supernova.  

An energy loss will affect the measured supernova distances of co-moving galaxies, shown in Fig.~1. 
\begin{figure}
\fbox{\includegraphics[width=5.in]{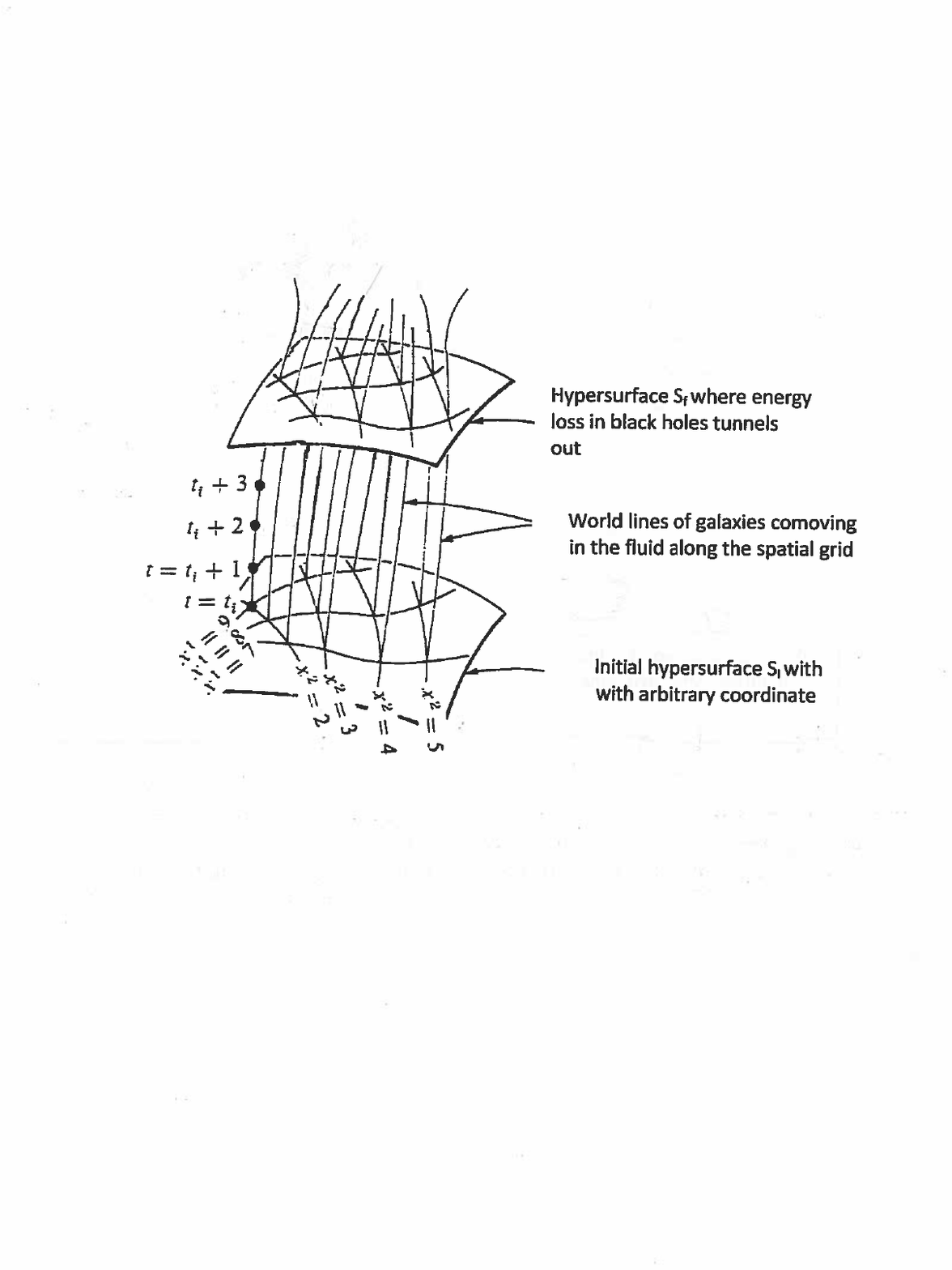}}
\caption{\textbf{Galaxy World Lines 3-Geometry}. This a a two-space and time
dimensional co-moving synchronous coordinate system for galaxies. Shown are two hyper-surfaces with arbitrary imposed grids. If the universe remained homogeneous and isotropic, then proper distances can be calculated. Due to central black hole energy losses, proper galactic distances are increased.}
\end{figure}
The 3-geometry $d\sigma^2= g_{ij}(t,x^t)dx^idx^j$ of each hyperspace is assumed to be the same
due to homogeneity of the universe. The initial hyper-surface $S_I$ 3-geometry is $\gamma_{ij}x^K \equiv g_{ij}(t_I,x^K)$. At time $t_I$ on surface $S_I$, they are separated by
the proper distance 
\begin{equation}
\Delta\sigma(t_I)=(\gamma_i,\Delta x^i \Delta x^j)^{1/2}. 
\end{equation}
At some later time
$t_f$, they will be separated by some other proper distance $\Delta\sigma(t_f)$. When spacetime is
isotropic, then the ratio $\Delta\sigma(t_f)/\Delta\sigma_(t_I)$ will be related to the universe scale factor $a(t_f)$ at time $t_f$. 
The Wilkinson Microwave Anisotropy probe combined with the Hubble Space Telescope resulted
in a very small value for the cosmological constant $\Lambda=3.73\times 10^{-56}cm^{-2}$ which corresponds to 
$\Omega_{\Lambda}=(\Lambda)/(3H_{0}^{2})=0.721\pm .015$\cite{Komatsu et al. (2011)}. Here $H_{0}=70.1 \pm 1.3 \, km/sec/Mpc$.
The loss in gravitational energy will cause galaxies no longer to be co-moving and to move away from the Hubble flow. The greater distances measured between them is known as dark energy. 

A limiting density with energy loss will solve the conflict in black holes between general relativity
and quantum theory. The conundrum is "Blacks holes: Complementary or Firewalls?"\cite{Almheiri et al. (2012)}. 
The smallest black hole $4M_{\odot}$, has the temperature of the black body Hawking radiation 
\begin{equation}
T=(\hbar c^{3})/(8\pi GMk_{B})=1.54\times 10^{-8}K 
\end{equation}
and power dissipation 
\begin{equation}
P = (\hbar c^{6})/(15360\pi G^{2}M^{2})=5.63\times 10^{-30}Watts
\end{equation}
This minuscule energy is maximal for a newly formed $4M_{\odot}$ black hole and will diminish further as it ages without accretion. 
Thus an infalling observer encounters nothing unusual at the horizon. Information will be released to the outside observer only when 
the black hole loses all its gravitational energy due to squeezing effects. No information encoding in Hawking radiation is necessary.

\section{A Cyclical Universe} 
 The Friedmann equation assumes a perfect fluid. It will be valid only hours into the big bang when large shell fragments captured hot core gases forming protogalaxies.  
The equations describing the scale factor evolution originate in the Ricci tensor. The 0-0 component of the Einstein equation is the Friedmann equation needing energy changes. In a Friedmann universe, radiation density is inversely related to the fourth power of 
the scale factor $\rho_{r} \propto a^{-4}$ and 
matter follows the third power $\rho_{m} \propto a^{-3}$.  
\begin{equation}
\Bigl(a_{,t}/a\Bigr)^{2}=-k/a^{2}+\Lambda/3+\, 8\pi\rho_{m}\,a_{o}^{3}/3a^{3}+\,8\pi\rho_{r}\,a_{o}^{4}/3a^{4}
\end{equation}
Here $a_{o}$ is our present day universe and $\Lambda$ is Einstein's term for energy of empty space. The question has long been why the initial geometry of the universe was flat or $k=0$ in the 
the above equation. If all the initial radiation $\rho_{r} $ was embedded in the matter and participated in the subsequent expansion and gravitation of a fairly large mass $r_{radius}\sim 10^{11}$ cm., then the flatness problem is explained. Due to the matching of gravitation and expansion energies ($k=0$ above), it is most unlikely anything but nucleons and leptons stopped the prior universe collapse or restarted the expansion. There was no free radiation prior to the big bang. The energetic core photons participated in the re-expansion and nucleosynthesis. The small dark energy value attributed to $\Lambda$ is actually due to later galactic black hole gravitational loss as explained.
Our universe cycles as follows as shown in Fig.~2.
\begin{figure}
\setlength{\fboxrule}{1pt}
\fbox{\includegraphics[width=5in,angle=0]{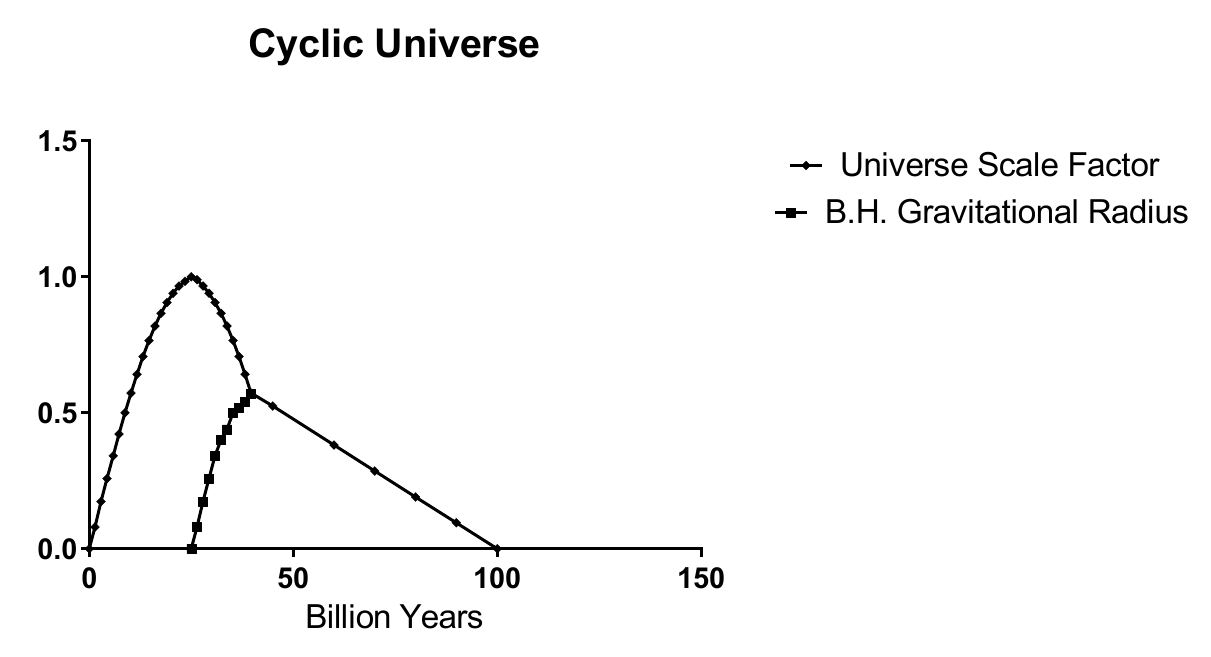}}
\caption{\textbf{The Cyclical Universe} started from a spherical shaped mass at or near limiting density. The universe expands to a maximum and then contracts. During contraction, there is a growing super-max black hole.  The collapsing universe will eventually force all matter and radiation into its gravitational radius. Kinetic energy of shell particles and radiation will slowly transfer to core. The gravitational energy and gravitational radius will slowly decrease to zero (cycle time estimated). Then core potential energy can start a new cycle of same size. Entropy is not increased.}
\end{figure}
During the universe contraction phase, there is a growing super-max black hole. All matter and photons inside its gravitational radius ($r_{+}\rightarrow 10^{26}$cm.) will follow null geodesics into the inside matter. Their energy will be slowly transferred to highly squeezed core particles. The collapsing universe will eventually force all remaining matter and radiation into the growing $r_{+}$. With the loss of gravitational energy and entropy, a subsequent universe bounce will not increase in size from the previous cycle. By bounce time, the gravitational radius has decreased to zero, $r_{+}\rightarrow 0$.
The very energetic core electrons and nucleons, loaded with radiation, powered the re-expansion as well as resulting gravity and nucleosynthesis.  
Nucleosynthesis thus occurred only in the core, leaving the shell as cold dark matter. 
If all the matter in the universe were in one super-mass, its radius would be $\sim 10^{12} cm.$ Its $4-5\%$ core released photons with an initial temperature $\sim 10^{17}$ degrees in a Planck spectrum. 
The cold baryonic shell surrounding the hot core absorbed and did not reflect hot core photons. 
It comprised a cavity close to the characteristic of a perfect black body with resulting 
radiation in thermal equilibrium as shown in Fig.~3.
\begin{figure}
\setlength{\fboxrule}{1pt}
\fbox{\includegraphics[width=.9\linewidth, angle=0.]{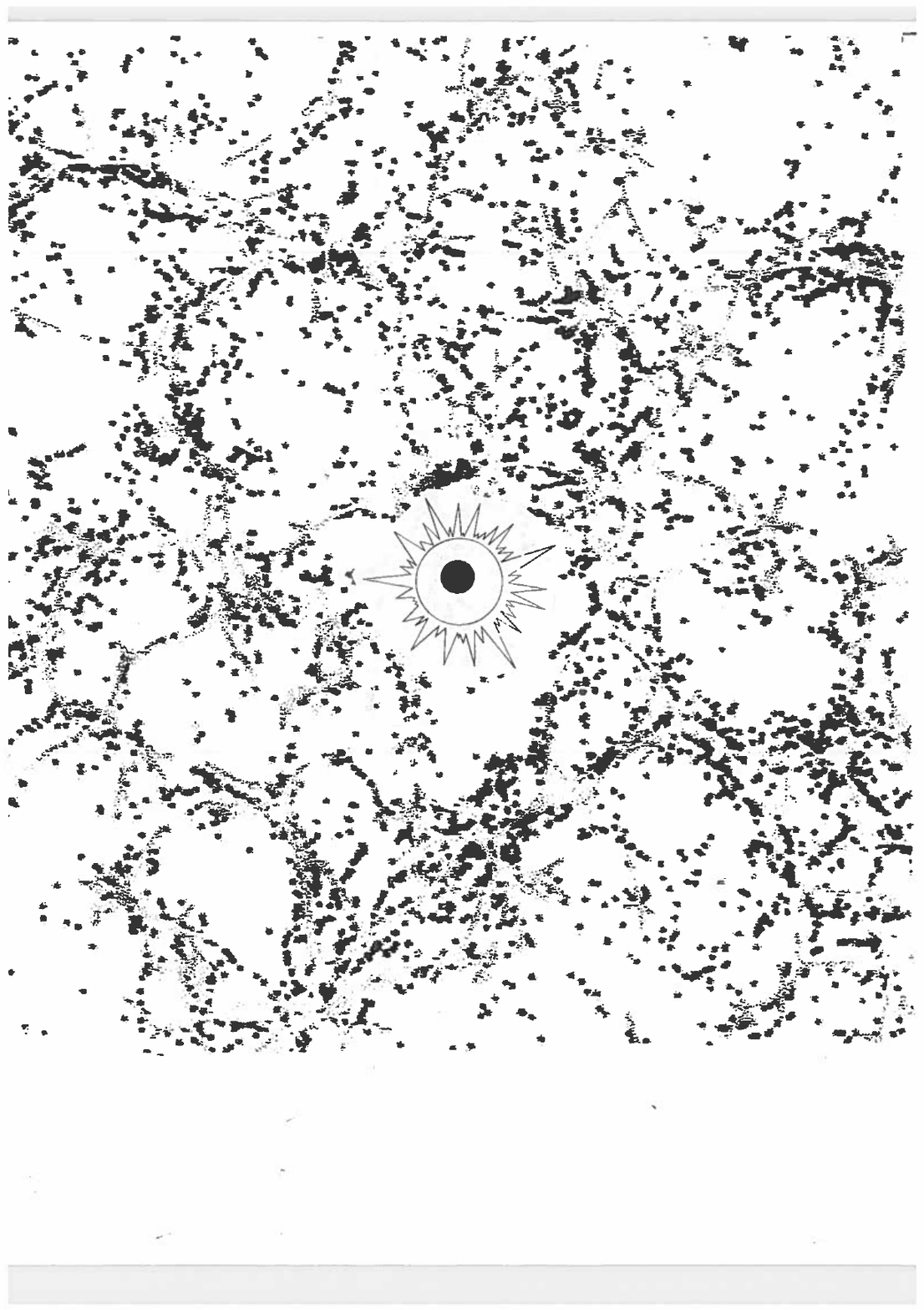}}
\caption{\textbf{Initial Mass Expansion.} A cold shell surrounds the hot core. After energy had been transferred to confined core baryons, gravitation loss permitted the expansion. The  core originated Planck spectrum radiation and shell originated dark matter and black holes.} 
\end{figure} 
 Whether released from a small hole or the massive break 
up of the shell, subsequent light emission would have a Planck spectrum with temperature fluctuations $\sim 10^{-5}$. 
The energy density was \cite{Brehm and Mullin (1989)}
\begin{equation}
u_\nu (T)= \Bigr( 8\pi\nu^2/c^3\Bigl) \Bigr(h\nu/(e^{h\nu/k_BT}-1)\Bigl)
\end{equation}
where $k_B$ is Boltzmann's constant. The first term on the right represents the number of electromagnetic modes of the standing waves at frequency $\nu$ per volume of cavity. The second term represents the average energy per mode at this frequency. The primordial spectrum of curvature perturbations can be represented as a power spectrum
\begin{equation}
\mathcal{P}(k) \propto k^{n_s -1}
\end{equation}
WMAP\cite{Komatsu et al. (2011)} showed that the power spectrum $n_{s}= 0.963 \pm .014$, which is nearly scale free.
The heights of the acoustic peaks are related to the related to the densities of the hot
and cold (CDM) baryons. The bulk modulus is reduced by increasing baryon fraction which
adds inertia but not pressure to the plasma. In the middle of the oscillations, the over-density
increases the compression peaks $(1,3,5 \ldots \,\,)$. The measured $\Omega_{b}h^{2}=0.02258 \pm .00056$
is consistent with nucleosynthesis. The angular scale of the acoustic peaks are related
to $r_{s}/d_{a}$, which is the sound horizon co-moving distance divided by the angular distance back to
the last scattering. This allows the spacetime curvature from the first acoustic peak to be measured. 
It was found to be flat to within $0.5\%$, all consistent with a bounce model. 

A 78 megahertz radio signal in the sky was caused by an unexpectedly strong interaction of primordial hydrogen with cold dark matter \cite{Barkana (2018)}. This signal in the cosmic radio frequency spectrum measuring up to $-600 mK$ (milliKelvins) is a clear sign of scattering of baryons with cold dark matter particles. With just baryon-cold dark matter cooling the signal could not be more than $-200 mK$. Its observed brightness related to the CMB in mK 
\begin{equation}
T_{21}\approx 0.023\,K\,\times x_{HI}(z) \Bigl[\Bigl(\frac{0.15}{\Omega_{m}}\Bigr)\Bigl(\frac{z+1}{10}\Bigr)\Bigr]^{0.5} \Bigl(\frac{\Omega_{b}}{0.02}\Bigr)\Bigl(1-\frac{T_{R}(z)}{T_{S}}\Bigr)
\end{equation}
$x_{HI}$ is the mass fraction of neutral non-ionized hydrogen. $\Omega_{m}$ is the cosmic mean matter density and $\Omega_{b}$ is the baryon part. $T_{S}$ is the 21 cm. spin-flip transition of hydrogen at z. $T_{R}$ is background radiation temperature.
The brightness temperature will be much enhanced from $-200mK$ in prior models to $-600mK$ if the cold dark matter is also neutral hydrogen.  
The cross-section for baryon-dark matter collisions $\sigma_{c}$. $\sigma_{1}$ is the cross-section assuming the relative velocity of $1$ km/sec. The dependence of the cross-section on velocity is assumed $v^{-4}$. The cross section increases greatly with decreasing velocity.
\begin{equation}
\sigma(v)=\sigma_{c}\Bigl(\frac{v}{c}\Bigr)^{-4}=\sigma_{1}\Bigl(\frac{v}{1 km/sec}\Bigr)^{-4}
\end{equation}
In order for the brightness temperature to be reached, a baryon size particle must be present. A brightness temperature of $-500 mK$ implies $\sigma_{1}>5 \times 10^{-21} cm^{2}$ and a cold dark matter particle $m_{\chi}< 1.5 GeV$.  $-300 mK$ implies a particle $m_{\chi} <4.3 GeV$. A particle $m_{\chi} <0.5 GeV$ would be unable to cool baryons sufficiently and make a maximum signal of log[$-T_{21}mK$] $2.2-2.3$ or $160-170 mk$ at best. With cold dark matter as atomic hydrogen $HI$, this enhancement is explained. The loss of Gaussian curve can be explained by  
early star formation in primordial galaxies producing small amounts of UV, X-ray or $\gamma$-ray radiation tending also to inhibit further star formation. Decoupling here is different from standard models as the hot gas is held in much lower orbits from most cold dark matter. A low gas temperature would enhance the Jeans mass necessary for early star formation. The dissipation of hot-cold baryonic relative velocities will also increase star formation. 
The initial capturing of energetic hot core baryons in lower black hole orbits with cold shell baryons in higher orbits will be explained below. Baryon inflo in dark matter halos was completed hours after the big bang.

\section{Current Problems in Galaxy Formation}

Large scale structure in the universe has long been assumed to grow from gravitational instabilities originating in primordial random Gaussian fluctuations. Galaxy growth then necessitates a number of essentially random stochastic processes involving mergers, accretion, gas expulsion and feedback from adjacent structures. Galaxies do congregate in clusters, along filaments and are absent from regions called voids. On the cluster and super cluster level, galaxies seem to have a somewhat random spacing. However on the galactic level, they are highly correlated in circular velocity, light emission, direction of rotation, etc.  

Current concepts in galaxy formation and growth are summarized 
as follows\cite{Loeb and Furlanetto (2013)}. A dark matter halo forms from random over-densities and acquires more matter through accretion or mergers. The gas then settles into a rotational supported disk. Within the disk-intrastellar medium(ISM), self regulation occurs through cooling and feedback. This causes ejection of a fraction of the disk's halo through winds. Gas flowing to the center of the galaxy may accrete onto a central black hole.  This unbinds other ISM gas through winds or jets. The halo number densities reflect the underlying dark matter with correlation function. Once the gas has cooled and collapsed to high density, star formation begins. 
Besides slow accretion, global instability such as bars, spiral waves and galaxy mergers can send gas to the super-massive black hole. Torques due to instabilities can send much gas to the galactic center. Galaxy evolution is critically dependent from feedback from stars and black hole. 

A review of the standard galaxy formation paradigm with $\Lambda CDM$ picture  adds some observation\cite{Pontzen and Governato (2014)}. There is a ubiquity of massive galactic outflows. These are driven by feedback phenomena such as ionizing radiation from young stars and black hole accretion as well as supernova winds. It also lists problems that still plague this paradigm.
The cusp-core discrepancy involves excessive dark matter at the center of dwarf galaxies in simulations.  
Higher mass galaxies with a peak rotational velocity $>100 km/s$ have flat central dark matter profiles. These have less inner distribution of dark matter than expected. Also dark matter and baryons follow similar profiles and may have tight coupling on a galactic scale. The angular momentum catastrophe in simulations has a large fraction of low angular momentum dark matter and baryons accumulating in all galactic halos. There is no understanding of the mechanisms that give a flat dark matter profile.
They comment that if sufficient energy could be given to dark matter particles in the center of the halo, they would migrate outward and the cold dark matter scenario could make normal galaxies.

Hot big bang models also suffer from the following problems. 

Galactic halo structures grew by attracting surrounding matter. The fate of these halos is determined either by radiative cooling or gravitational heating. In low mass halos, cooling predominates, which allows cold gas to fall into the center and become disks and stars\cite{Cattaneo et al. (2009)}. The cooling problem is most acute in galaxies. At the end of their lives, massive stars return $30-40\%$  of their mass to the ISM. If even a small
fraction of this mass is accreted, it would result in much larger black
holes than are present. Gravitational heating dominates once a halo mass of about $10^{12}M_\odot$ is reached. Cold gas is no longer able to accrete onto galaxies. The only way galaxies within halos can grow at this stage is by mergers. 

Pure disk galaxies form bulges after the mergers. Yet some samples of  giant galaxies have found over half are large pure disk type, without any evidence of mergers\cite{Kormendy et al. (2009)}. 

The intergalactic medium of the Perseus cluster was found to have a uniform
iron enrichment of about one third that of our sun
\cite{Werner et al. (2013)}. Uniform metal enrichment must have occurred prior to the formation of the cluster and its supernovas. 

In galaxies with bulges, the mass of the central black hole correlates with the mass of the bulge and also the average spread of velocities of the bulge stars. This includes ellipticals which have bulges but no disks
\cite{Peebles (2011)}. Why did the gas that formed bulge stars settle near the black hole? Part increased the black hole mass and part led to explosions that blew the gas away and suppressed star formation. What is the reason for the inward movement of matter to the black hole in some galaxies and to the pseudo-bulge in others? Accepted theory has both galaxies with and without bulges growing by accreting matter during the period that the massive early stars were  forming. These early stars would not have settled in disks because they could not be slowed enough to reside in disks. Galactic bulges do contain old stars but there is no reason these old stars avoided bulge-less galaxies. There is no evidence that they are in diffuse stellar halos either. Evidently they played little role in galaxy formation. 

The mass distribution of spiral galaxies is evenly spread from its dark matter outer limits to its inner baryonic areas. Dark matter played a strong role in the disk and stars but not its black hole. In pure disk galaxies with pseudo-bulges, the central black hole does not correlate with the pseudo-bulge\cite{Kormendy et al. (2011)}.  

There is extremely inefficient star formation in metal poor galaxies\cite{Shi et al. (2014)}. The universe's original  Pop III stars must have formed very slowly in the initially metal poor galaxies. They had to use their fuel and collapse to form initial black holes extremely rapidly.  

There is evidence that galactic size and brightness have not increased since $Z \approx 1.25$\cite{Lerner (2017)}.
Plots for many elliptical and the disk galaxies show the radius vs. $1+z$ in log-log plots yield a flat line. In otherwords there is no change of average galactic radius or brightness with time.

\section{Baryonic Galaxy Formation}

Galactic properties have been shown to be a function of only one variable: the central black hole mass\cite{Vandenbergh (2008)}. For most galaxies, there is a uniform history for galactic evolution\cite{Steinhardt (2014)}. There is a synchronization timescale $(\tau_{s}\approx 1.5 Gyr)$ where  galaxies of fixed mass and red-shift go through a deterministic sequence of star formation, quasar accretion and eventual quiescence. This sequence negates the importance of stochastic processes.
Protogalaxies must form in the very initial stages of the radiation dominant era. Radiation would not be able to destroy structures containing black holes with orbiting hot hydrogen and helium gases. The integrated Sachs-Wolf effect 
\begin{equation}
\ell^{2}C^{ISW}_{\ell} \simeq 72\pi^{2}/25\ell\int^{r_{LS}}_{0} dr\, r
\, g^{\prime 2}(r) \mathcal{P}_{\mathcal{R}}\Bigl(\ell/r\Bigr)T^{2}
\Bigl(\ell/r\Bigr)
\end{equation}
was designed for photons descending into and emerging from a gravitational well. Here $\mathcal{P}_{\mathcal{R}}(k)$ is the primordial co-moving curvature spectrum.
$r_{LS}$ is the co-moving radius to the last scattering surface. 
$g$ is the $\Lambda$ growth suppression factor. $T(k)$ is the transfer function for suppression during radiation domination. $k^{\prime}$ is the conformal time derivative of the co-moving wave number. Note it does not include effects of orbiting hot gases with electrons obscuring matter imprints on the CMBR photons. Thus structures much larger than the currently allowed 
$ 10^{4} M_\odot$ 
could have been present in the last scattering 
surface without enlarging the isotropy from $10^{-5}$. 
Starting from a big bang shell and hot core, early galaxy formation and high inter-galactic correlations become simplified. The shell laid down a fairly even cold dark matter density $\rho_{DM}$. 
The DM halo mass distribution for galactic systems ranging from dwarf discs
and spheroidals to spirals and ellipticals has been found essentially constant\cite{Donato et al. (2009)}. This result also spans almost the whole galaxy magnitude range $M_B$ from
$-8$ to $ -22$ and gaseous to stellar mass fraction range of many orders of magnitude.
\begin{equation}
log(\mu_{0D}/M_\odot pc^{-2})=\, 2.15 \pm 0.25
\end{equation}
where $\mu_{0D}$ is the central surface density and is defined as 
$r_0\rho_0$. $r_0$ is the halo core radius and $\rho_0$ is the central density. This same finding was supported by another group
\cite{Gentile et al. (2009)}.
Since all dark matter lies within a halo orbiting the primordial black holes, its total density field is
\begin{equation}
\rho(\mathbf{x})=\Sigma_{i} \int dm \int d^{3}x\prime \delta(m-m_{i})\delta(\mathbf{x}\prime-\mathbf{x}_{i})m\, u(\mathbf{x}-\mathbf{x}\prime|m)
\end{equation}
where $i$ is the different halos, $u=\rho/M$ is the normalized density profile and $M$  is the halo virial mass. 
Within the original dark matter halos, massive black holes coalesced. 
The dark matter from the center to periphery of early type galaxies has been evaluated from a galactic stellar mass $M_{*}\sim 10^{10}M_{\odot}$ to the more massive galaxies $10^{12}M_{\odot}$\cite{Tortora et al. (2014)}.
N-body simulations predict the the dark matter density profile $\rho_{DM}(r)$ should be independent of halo mass. In the NFW profile it is described by two power laws. In the outer regions it is $\rho_{DM}(r)\propto r^{-3}$ and in the center $\rho_{DM}(r)\propto r^{\alpha}$. Here $\alpha$ can vary $-1$ or $-1.5$ depending on the model. What has actually been found is a variation around the inverse gravitational square law ($\alpha=-2$), as the halos are orbiting the primordial black holes.
Galaxies larger than $ 3\times 10^{10}M_{*}$ have lower slopes than 
 $\alpha\approx 2$ due to accretion of halo mass from smaller satellite galaxies. Smaller satellite galaxies under $3 \times 10^{10}M_{*}$ have larger slopes due to loss of outlying halo matter.  
The rotation of the larger galaxies above this break point has been found  highly correlated and perpendicular to the filament that they are located
\cite{Dubois et al. (2014)}. The rotation of $65$ galactic black holes has been found aligned using their radio galactic jets \cite{Taylor and Jagannathan (2016)}.  The rotation of the smaller galaxies has been found parallel to the cosmic web that they belong as shown in Fig.~4. The hot core blast wave, which followed after the dark matter-black hole foundation, crossed these structures in a mostly perpendicular direction. The already formed larger black holes, with rotation already in the plane of their filaments, did not have their rotation direction changed during the capturing process. The smaller galaxies, in orbit around the larger galaxies as satellites, were forced  to change their rotation into the plane of the filament. 
\begin{figure}
\centering
\setlength{\fboxrule}{1pt}
\fbox{\includegraphics[width=0.9\linewidth]{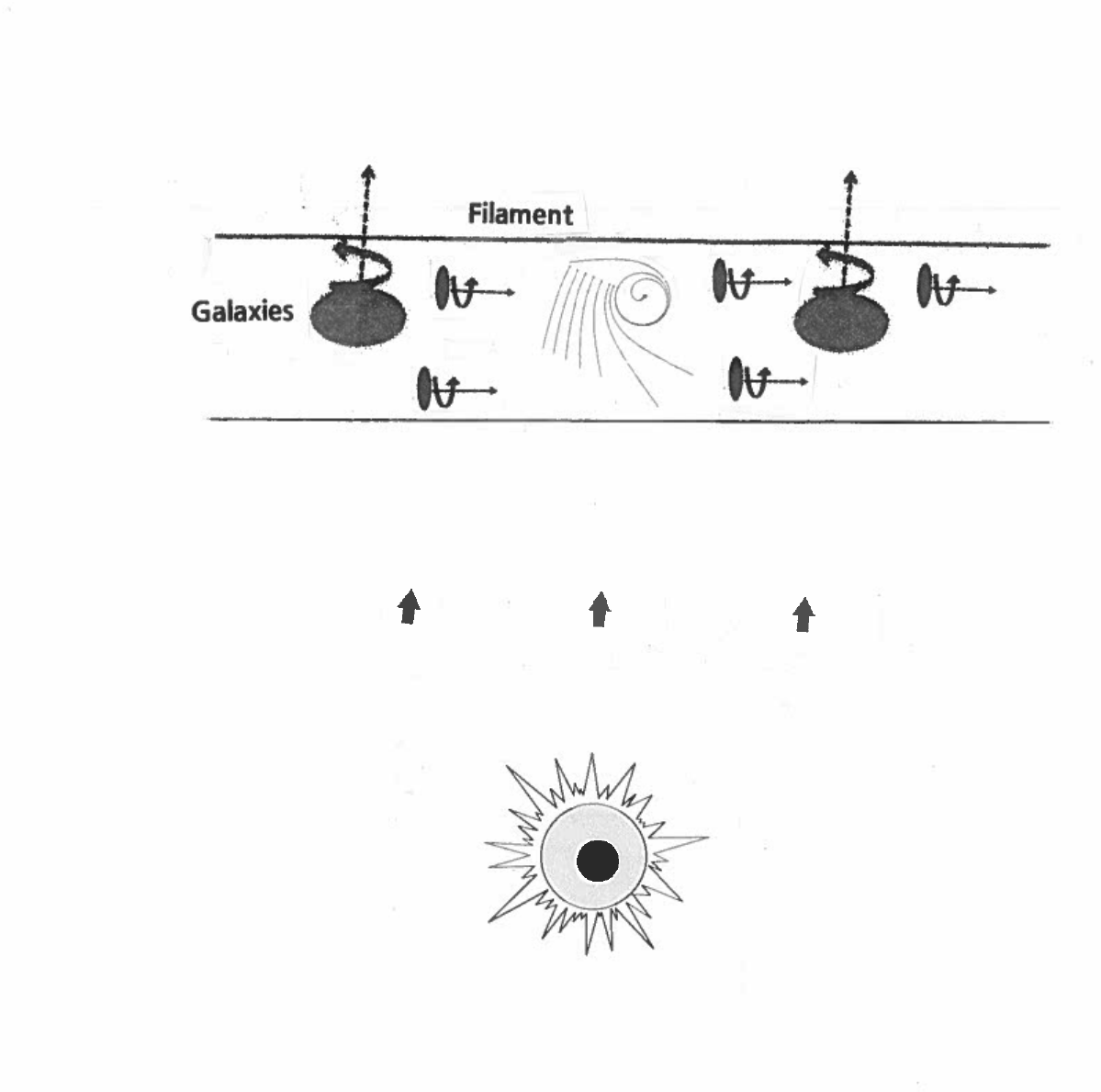}}
\caption{\textbf{ Early Galaxy Formation.} Large and small shell masses were driven into the universe forming filaments. Larger masses coalesced into black holes which held a fairly even density of smaller dark matter. Later hot core gas was captured according to the depth of the gravitational well. During capturing process, larger black holes, unlike smaller black holes, did not change the direction of rotation.}
\end{figure}
The deeper the gravitational 
wells, the higher the velocity and more orbiting mass that could 
be captured. 
The capturing process described here is divided by distance from the primordial black holes $M$.
Outside the immediate area of black hole influence, capturing of hot core matter $m$ streaming through the area of influence of each black hole is due to the amount of energy each particle possesses. Large
kinetic energies result in hyperbolic or parabolic type orbits with the ability to escape any given gravitational well. Matter that is captured has the potential energy greater than the kinetic, 
\begin{equation}
GmM/r>\, l^2/mr^2\,+1/2 \,m\dot{r}^2
\end{equation}
and $e<1$. Expanding the total kinetic energy $E$ in the equation for $e$,
\begin{equation}
e=\sqrt{1+(2l^2(l^2/mr^2+\,1/2\, m\dot{r}^{2}-GmM/r))/mk^{2}}
\end{equation}
Orbiting matter has $e<1$ and real. If we let its angular momentum $l=mr\dot \theta^2$ and $k=mMG$, the equation for $e$ becomes 
\begin{equation}
e=\sqrt{1+(r^6\dot{\theta}^{4}+\dot{r}^{2}r^{4}\dot{\theta}^2
-2GMr^{3}\dot{\theta}^{2})/(M^{2}G^{2}})
\end{equation}
Using $\dot\theta=\dot r/r$, the equation for $e$ becomes
\begin{equation}
e=\sqrt{1+(2r^{2}\dot{r}^{4})/(M^2G^2)-(2r\dot{r}^{2})/(MG)}
\end{equation}
As $GM=\dot r^2r$, 
then the galactic well will deepen as $M\propto \dot r^2$
or $M\propto r$. The last term in equation above becomes $\dot{r}^{8}/M^{2}G^{2}$. When this
term is dominant, it will allow capturing matter with $\dot{r}$ to 
increase as the fourth power as the galactic black hole $M$ increases,
$\dot{r} \propto M^{4}$. This explains the Tully-Fisher and similar correlations\cite{Shaya and Tully (2013)}.
The black hole capturing cross sectional area, $M_{csa}\propto M_{gravity}$ since both scale as $r^{2}$. 

With the two stage gravitational formation process, angular momentum is preserved. Halo parameters are related to the luminous mass distribution since all rotating mass was captured by a given size black hole.  
An entirely baryonic model explains why the circular orbital speed 
from luminous matter, which dominates the inner regions, is so similar
to dark matter at larger radii. With many stars in the center areas, initial
conditions for dark and luminous matter no longer have to be closely adjusted 
to produce a flat rotation curve\cite{Ibata et al. (2013)}. 
The hot core matter of a certain velocity can be captured by the similarly sized black holes, 
explaining why there are similar circular speeds in 
all galaxies of a given luminosity no matter how the luminous matter is spaced. The depth of the gravitational well determines the circular speed
and luminosity of captured matter. The hot and cold matter discrepancies are detectable only at
accelerations below $\sim 10^{-8}cm/sec^{2}$ since they are all baryons. Much of the missing bayonic matter has been found in the intergalactic medium\cite{Nicastro (2018)}.

\section{Discussion}
General relativity has been extrapolated in black holes and the big bang to 
enormous energies and densities without consideration of loss of space in particles.   
Using evidence of huge resistive pressures in the stress-energy tensor $T_{\mu\nu}\rightarrow \mathbf{0}$, very simple 'quantum gravity' can be produced. The gravitational loss in neutron stars can't be explained any other way. General relativity remains valid throughout the universe, once loss of space in nucleons is included. This caused the initial in-homogeneous big bang configuration of matter. A cold shell and hot core produced
the Planck spectrum radiation. The cold shell dispersion led to the great wall, filaments and voids.
The basis of all galaxies formed simultaneously as cold dark shell matter formed black holes which captured subsequent hot core
gases into protogalaxies. These formed stars ionized the intergalactic medium from $z=6-11$. The fact that the Higgs Boson is $125$ GeV and not larger, eliminates nonbaryonic matter as dark matter. 
The pre-big bang matter was composed of highly squeezed ordinary nucleons. It was definitely not a plasma or a singularity.

Gravitation is described by the Riemann metric.  No negative energies are required nor is a vacuum energy necessary for empty space.
The fact that initial universe had flat space time despite the apparent early radiation energy dominance has been explained. 
Since the origin of the big bang is a bounce, it should recollapse into the big crunch $\sim 100$ billion years. The resulting neutrons will eventually decompose into the protons and electrons for hydrogen and helium during the next cycle.   
Black holes must continuously accrete mass-energy to maintain 
their gravitational strength. Gravity is not caused by matter itself but rather by the motion of matter particles.  Galaxies are made in a two stage process.  First the big bang shell laid down all halos and super-massive black holes. Then hot core gases were captured according to the size of each gravitational well.
\twocolumn

\bibliography{galaxy4d.tex}
\tiny
\textbf{Acknowledgments} This article is based on research I posted on the internet astrophysics archives while in NJIT Physics. Some of it was done in collaboration with 
John Rollino of Rutgers University Physics, Newark N.J. 07102 I wants to thank Erik Schnetter for helpful advice.
 
\end{document}